\documentclass[aps,prb,twocolumn,showpacs,nofootinbib,groupedaddress,amsfonts,amssymb,amsmath]{revtex4-1}
\usepackage{graphicx}
\usepackage{epstopdf}
\usepackage[colorlinks=true, letterpaper=true, pdfstartview=FitV, linkcolor=blue, citecolor=blue, urlcolor=blue]{hyperref}
\usepackage{appendix}

\begin{document}

\title{Quantum nonlinear ac transport theory at low frequency}

\author{Lei Zhang,$^{1,5,\dagger}$ Fuming Xu,$^{2,\dagger}$ Jian Chen,$^{3}$ Yanxia Xing,$^{4}$ Jian Wang$^{2,3,*}$}


\address{$^1$State Key Laboratory of Quantum Optics and Quantum Optics Devices, Institute of Laser Spectroscopy, Shanxi University, Taiyuan 030006, China\\
$^2$College of Physics and Optoelectronic Engineering, Shenzhen University, Shenzhen 518060, China\\
$^3$Department of Physics, The University of Hong Kong, Pokfulam Road, Hong Kong, China\\
$^4$Beijing Key Laboratory of Nanophotonics and Ultrafine Optoelectronic Systems, Beijing Institute of Technology, Beijing 100081, China\\
$^5$Collaborative Innovation Center of Extreme Optics, Shanxi University, Taiyuan 030006, China}

\begin{abstract}
Based on the nonequilibrium Green's function (NEGF), we develop a quantum nonlinear theory to study time-dependent ac transport properties in the low frequency and nonlinear bias voltage regimes. By expanding NEGF in terms of time to the linear order in Wigner representation, we can explicitly include the time-dependent self-consistent Coulomb interaction induced by external ac bias. Hence this theory automatically satisfies two basic requirements, i.e., current conservation and gauge invariance. Within this theory, the nonlinear ac current can be evaluated at arbitrarily large bias voltages under the low frequency limit. In addition, we obtain the expression of time-dependent current under the wide band limit and derive the relation between the nonlinear electrochemical capacitance and the bias voltage, which are very useful in predicting the dynamical properties of nanoelectronic devices. This quantum theory can be directly combined with density functional theory to investigate time-dependent ac transport from first-principles calculation.
\end{abstract}

\maketitle

\def\thefootnote{*}\footnotetext{These authors contributed equally to this work}

\section{introduction}

Nonlinear Hall effect\cite{L-Fu,Guinea1,NHEREV1,NHEREV2} is the emerging frontier of condensed matter research\cite{Low,Sodemann,G-Su,Tewari2,HJiang2,W-Zhang,S-Zhang,Saha,MWei,MWei2022,Narayan,Ueda,Yarnjp,Wei2023}. Prominent nonlinear Hall phenomena include the second order Hall effect discovered in WTe$_2$\cite{S-Xu,Q-Ma,K-Kang} as well as TaIrTe$_4$\cite{H-Yang}, and the third order Hall effect verified in T$_d$-MoTe$_2$\cite{S-Lai}, etc. These higher-order responses are relatively weak and their measurement requires the phase lock-in amplifier technique, where low frequency ac bias voltage is applied as the driving signal while the second \cite{Q-Ma,K-Kang,H-Yang} and third harmonic\cite{S-Lai} responses labeling nonlinear signals are probed. However, theoretical interpretation of nonlinear Hall effects is mostly based on the semiclassical Boltzmann approach. In terms of the second \cite{MWei} and third order\cite{MWei2022} conductances expressed in nonequilibrium Green's function (NEGF), quantum transport properties of nonlinear Hall effects have been discussed, but only in the dc case. Therefore, a nonlinear ac transport theory at low frequency is on demand and definitely helps to comprehend nonlinear Hall effects.

On the other hand, time-dependent quantum transport in molecular and nanoscale devices have attracted intensive research attention both experimentally\cite{Feve1,Feve2,Gabelli,McEuen,Flindt,Haug,Price,Umbach,Howard,Shayegan,Harmans} and theoretically\cite{Ting,Pastawski,buttiker0,Wingreen,Datta,Schoeller,Ng,Ivanov,jauho1,bgwang,bin0,Gross,Maciejko,buttiker1,buttiker2,Ferry,jian0,
mackinnon,Waintal,Xing,lei1,lei2,bin,xufm,gaomin,Yu,Kienle,Waintal0,Max,tao,Korniyenko,lei2019}. Time-dependent response in quantum transport is of great importance, since it can provide critical information on dynamical properties that are absent in the dc case. The dynamical property is essential in accurate design of nanoelectronic devices. Under time-dependent external fields, quantum transport investigation focuses on two different regimes, i.e., the transient regime when the external bias is immediately turn on or off and the long time ac steady-state regime. For the transient dynamics that characterizes the relaxation time of an electronic system when turned on or off, various approaches have been developed, including scattering matrix theory\cite{buttiker0,buttiker1,buttiker2}, NEGF\cite{Wingreen,jauho,Maciejko}, and time-dependent Schr\"{o}dinger equation\cite{Gross}. In the ac steady-state regime, B\"{u}ttiker \emph{et al} derived a current conservation theory which explicitly includes the displacement current in terms of scattering matrix\cite{buttiker0}. However, this theory is only applicable under near-equilibrium condition. Later, the situation has been extended to the far-from-equilibrium condition by using NEGF\cite{ydw2}.

According to Ref. [\onlinecite{buttiker0}], the long-range Coulomb interaction must be included in time-dependent and nonlinear dc quantum transport theories, which is necessary for satisfying two basic requirements, i.e., current conservation and gauge invariance\cite{buttiker0,lei3}. The Coulomb interaction can be treated at least in the Hartree level in density functional theory (DFT). Therefore, it is convenient to adopt the first-principles approach, i.e., DFT carried out within the Keldysh NEGF (NEGF-DFT) formalism, to study the transport properties of nanoelectronic devices. In practice, quantitative predictions of their transport properties were compared with experimental results\cite{Kaun1,Kaun2,Frederiksen,Reed}. Although the NEGF-DFT formalism has been widely applied in predicting dc quantum transport properties, its applications to ac situations received far less attention. This is due to the lack of a quantum ac transport theory in the nonlinear bias voltage regime. Such an ac transport theory can be directly coupled with DFT to predict quantum nonlinear ac transport properties from atomic first principles, which is absent so far. Therefore, a nonlinear ac NEGF theory is urgently needed.

In this work, we develop a quantum nonlinear formalism to study ac transport properties at low frequency, where Coulomb interaction is explicitly treated within the NEGF formalism. Based on the adiabatic approximation, we expand the Floquet NEGF with respect to Frozen Green's function in Wigner representation at low frequency. Physically, Frozen Green's function describes that the potential experienced by electrons during their transport is instantaneously adjusted to the applied ac bias. Following this route, the nonlinear time-dependent current is obtained. Furthermore, a concise formula is given under the wide band limit (WBL), which can recover the dynamic conductance at small ac bias\cite{ydw2}. In particular, we demonstrate that the present theory can be used to study the nonlinear electrochemical capacitance in response to external bias, i.e., the C-V curve.

The paper is organized as follows. In Sec.{\color{blue}II}, we introduce the theoretical model for ac transport. Sec.{\color{blue}III} presents the derivation of time-dependent NEGF in terms of Frozen Green's function. In Sec.{\color{blue}IV}, we provide the time-dependent ac current formula and discuss two limiting cases, i.e., wide band limit and small ac bias limit. Nonlinear electrochemical capacitance is also discussed. Finally, a brief summary is given in Sec.{\color{blue}V}.

\section{Model}
\begin{figure}
\includegraphics[width=8cm,totalheight=2.5cm]{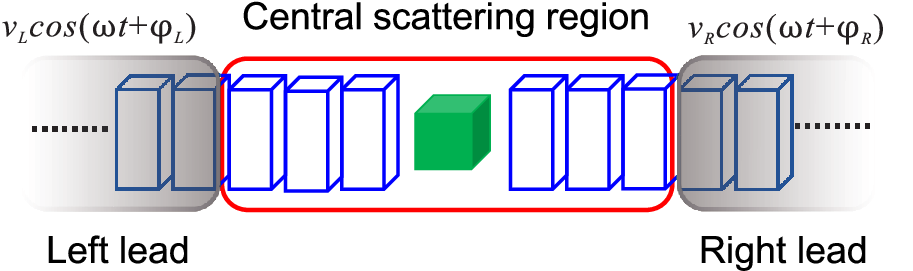}
\caption{Schematic plot of a two-terminal nanoelectronic device consisting of a central scattering region (red line box) and two leads. The left and right leads extend to electron reservoirs at infinity, where time-dependent bias voltages $v_{\alpha}cos(\omega t+\varphi_{\alpha})$ are applied and electric currents are probed.} \label{Fig1}
\end{figure}
In this section, we introduce the model system for ac transport. The system under investigation is a quantum device in contact with two leads extending to electron reservoirs where time-dependent ac bias is applied, as shown in Fig.~\ref{Fig1}. The Hamiltonian of this model system can be expressed in three parts
\begin{eqnarray}
H = H_{\alpha} + H_{C} + H_T.
\end{eqnarray}
Here $H_{\alpha}$ is the Hamiltonian of lead $\alpha$
\begin{eqnarray}
H_{\alpha} = \sum_{k\alpha} \epsilon_{k\alpha}(t) C^{\dagger}_{k\alpha} C_{k\alpha},
\end{eqnarray}
where $C^\dagger_{k\alpha}/C_{k\alpha}$ creates/annihilates an electron in lead $\alpha$. $\epsilon_{k\alpha}(t)=\epsilon^{(0)}_{k\alpha} + qv_\alpha(t)$, where $\epsilon^{(0)}_{k\alpha}$ is the energy level of lead $\alpha$ and $v_\alpha(t)=v_{\alpha}\mathrm{cos}(\omega t+\varphi_{\alpha})$ is the time-dependent external ac bias. $\omega$ is the frequency of this ac bias and $\varphi_{\alpha}$ denotes the corresponding phase. The second term $H_{C}$ is the Hamiltonian of the isolated central region,
\begin{eqnarray}
H_{C}= \sum_n (\epsilon_n + q U_n) d^\dagger_n d_n,
\end{eqnarray}
where $d_n^\dagger/d_n$ is the creation/annihilation operator of the electron in the central scattering region. Internal Coulomb potential $U_n$ under the Hartree approximation is included in the central region, which is defined as
\begin{equation}
U_n = \sum_m V_{nm} \langle d^\dagger_{m} d_{m}\rangle,
\label{X2coulomb}
\end{equation}
where $V_{nm}$ is the matrix element of the Coulomb potential. In real space representation, $V(x,x') = q/|x-x'|$. The exchange and correlation interactions can be treated in a similar way. The last term $H_T$ describes the coupling between the central region and leads
\begin{eqnarray}
H_T=\sum_{k\alpha n}[t_{k\alpha n} C^{\dagger}_{k\alpha} d_n
+t^*_{k\alpha n} d^\dagger_n C_{k\alpha} ],
\end{eqnarray}
with $t_{k\alpha n}$ the coupling constant.

Based on NEGF, the time-dependent charge current flowing in lead $\alpha$ can be written as ($\hbar=-q=1$)\cite{jauho1},
\begin{equation}
\begin{split}
I_\alpha(t) =& \int dt_1 {\rm Tr} [G^r(t,t_1) \Sigma^<_{\alpha}(t_1,t) + G^<(t,t_1) \Sigma^a_{\alpha}(t_1,t)\\
&\qquad - \Sigma^<_{\alpha}(t,t_1) G^a(t_1,t)- \Sigma^r_{\alpha}(t,t_1) G^<(t_1,t)].\label{current-1}
\end{split}
\end{equation}
Here $G^{r,a,<}$ ($\Sigma^{r,a,<}$) is the time-dependent Green's function (self energy) with double time indices in principle. In order to obtain the time-dependent current $I_{\alpha}(t)$, one has to know $G^{r,a,<}$ and $\Sigma^{r,a,<}$. In the following section, we will discuss how to calculate them in Wigner representation by expanding them to the first order in frequency, and then evaluate $I_{\alpha}(t)$.

\section{Green's function and self energy in Wigner representation}

We start from the time-dependent retarded and lesser Green's functions of the central scattering region on real-time axis\cite{jauho},
\begin{equation}
\begin{split}
[i\frac{\partial}{\partial t}-H_C(t)]G^r(t,t')&=\delta(t-t')\\
&+\int\Sigma^r(t,t_1)G^r(t_1,t')dt_1,\label{eq-Gr-1}
\end{split}
\end{equation}
and
\begin{equation}
\begin{split}
[i\frac{\partial}{\partial t}-H_C(t)]G^{<}(t,t')& =\int dt_1 [\Sigma^r(t,t_1)G^{<}(t_1,t') \\
&+ \Sigma^{<}(t,t_1)G^{a}(t_1,t')].\label{eq-Gl-1}
\end{split}
\end{equation}
In general, the Hamiltonian $H_C(t)$ in Eqs.~(\ref{eq-Gr-1}) and (\ref{eq-Gl-1}) can be separated into two parts, i.e., $H_C(t)=H_{C0}+U(t)$, where $H_{C0}$ is the time-independent part and $U(t)$ is the time-dependent internal Coulomb potential in response to the external ac bias.  To develop a formalism that can be combined with DFT, the Coulomb potential $U(t)$ should be solved explicitly through the Poisson equation
\begin{equation}
\begin{split}
\nabla^2U(x,t)= -4\pi\rho(x,t) = -4\pi i [G^<(t,t)]_{xx},\label{eq-Poisson-1}
\end{split}
\end{equation}
where $\rho(x,t)$ is the charge density and $x$ represents the real space position in the central region. Note that Eqs.~(\ref{eq-Gr-1}), (\ref{eq-Gl-1}), and (\ref{eq-Poisson-1}) are coupled, which means that they should be solved self-consistently.

The Green's functions in Eqs.~(\ref{eq-Gr-1}) and (\ref{eq-Gl-1}) depend on two time indices $(t,t^{'})$, since the time translational invariance is broken due to the presence of the time-dependent ac bias voltage applied in the lead. In the following, we first express the two-time function $A(t_1,t_2)$ in Wigner coordinates, where a slow classical timescale $t=(t_1+t_2)/2$ and a fast quantum timescale $\tau=t_1-t_2$ are introduced\cite{jauho}. Thus, $A(t_1,t_2)$ is transformed into $A(t,\tau)$ and an integral expression such as
\begin{equation}
A(t_1,t_2) = \int dt_3 A_1(t_1,t_3)A_2(t_3,t_2),
\end{equation}
gives \cite{jauho}
\begin{equation}
A(t,E) = A_1(t,E)\hat{{\mathcal F}}(t,E)A_2(t,E),
\end{equation}
where $A(t,E) = \int d\tau e^{iE\tau}A(t,\tau)$ and
\begin{equation}
\hat{{\mathcal F}}(t,E) = \exp\Biggl(-\frac{i}{2}\biggl[\frac{\overleftarrow{\partial}}{\partial t}\frac{\overrightarrow{\partial}}{\partial E} - \frac{\overleftarrow{\partial}}{\partial E}\frac{\overrightarrow{\partial}}{\partial t}\biggr]\Biggr),
\label{gradient-op}
\end{equation}
is the gradient operator and arrow indicates the direction of differentiation. When the external ac bias is slow enough compared to characteristic timescales of the electron in the system, the gradient operator in Eq.~(\ref{gradient-op}) can be expanded order by order in time or energy. Here, we keep only the first order of the Taylor expansion
\begin{equation}
\hat{{\mathcal F}}(t,E) \approx 1 - \frac{i}{2}\biggl(\frac{\overleftarrow{\partial}}{\partial t}\frac{\overrightarrow{\partial}}{\partial E} - \frac{\overleftarrow{\partial}}{\partial E}\frac{\overrightarrow{\partial}}{\partial t}\biggr). \label{gradient-op-1}
\end{equation}
For simplicity, abbreviations for derivatives $\frac{\partial}{\partial E}=\partial_E$ and $\frac{\partial}{\partial t}=\partial_t$ are adopted in the following derivation. Thus Wigner transformation of a two-time function gives\cite{jauho}
\begin{equation}
A(t,E) = A_1(t,E)A_2(t,E) + \frac{i}{2}\{A_1,A_2\},\label{eqAB1}
\end{equation}
where Poisson bracket is defined as
\begin{equation}
\{A_1(t,E),A_2(t,E)\} = \partial_E A_1\partial_t A_2 - \partial_t A_1\partial_E A_2.\label{bracket}
\end{equation}

Therefore, by taking Wigner transformation and using Eq.~(\ref{eqAB1}), the corresponding Floquet Green's functions of Eqs.~(\ref{eq-Gr-1}) and (\ref{eq-Gl-1}) up to the first order become (detailed derivation is presented in Appendix {\color{blue}A})
\begin{equation}
\begin{split}
&\frac{i}{2}\partial_tG^r(t,E)+\frac{i}{2}\partial_tH_C(t)\partial_EG^r(t,E)\\
&+[E-H_C(t)-\Sigma^r(t,E)]G^r(t,E) -\frac{i}{2}\left\{\Sigma^r,G^r\right\}=I,\label{eq-Gr-2}
\end{split}
\end{equation}
and
\begin{equation}
\begin{split}
&\frac{i}{2}\partial_tG^{<}(t,E)+\frac{i}{2}\partial_tH_C(t)\partial_EG^{<}(t,E)\\
&+[E-H_C(t)-\Sigma^r(t,E)]G^{<}(t,E) -\Sigma^<(t,E)G^{a}(t,E)\\
&-\frac{i}{2}[\left\{\Sigma^r,G^<\right\}+\left\{\Sigma^<,G^a\right\}]=0.\label{eq-Gl-2}
\end{split}
\end{equation}
Here we have used $C(t_1,t_2)=A(t_1)B(t_1,t_2)=\int dt_3 A(t_1,t_3)B(t_3,t_2)\delta(t_1-t_3)$ and its Wigner transformation is expressed as
\begin{equation}
\begin{split}
A(t_1)B(t_1,t_2)\Longrightarrow A(t)B(t,E)-\frac{i}{2}\partial_t A \partial_E B.\label{eqAB-1}
\end{split}
\end{equation}

In order to calculate the Green's function $G^{r/<}(t,E)$, the self energy $\Sigma^{r/<}(t,E)$ should be expanded up to the first order in time. We show that the self energy up to the first order is the same as that of the equilibrium one.

We start from the retarded Green's function $g^r_{\alpha}(t,E)$ of lead $\alpha$ first, since the self energy $\Sigma^r_{\alpha}(t,E)$ is defined as $\Sigma^r_{\alpha}(t,E)=H_{C\alpha}g^r_{\alpha}(t,E)H_{\alpha C}$ with $H_{C\alpha}$ describing the coupling between the central region and lead $\alpha$. Similar to Eq.~(\ref{eq-Gr-2}), the retarded Green's function of lead $\alpha$ can be expressed as
\begin{equation}
\begin{split}
\frac{i}{2}\partial_tg^r_{\alpha}(t,E)&+\frac{i}{2}\partial_tH_{\alpha}(t)\partial_Eg^r_{\alpha}(t,E)\\
&+[E+i\eta-H_{\alpha\alpha}(t)]g^r_{\alpha}(t,E)=I,\label{eq-LGr-1}
\end{split}
\end{equation}
where $\eta$ is an infinity small positive number. Let's introduce the Frozen Green's function of the lead,
\begin{equation}
\begin{split}
g^r_{\alpha,f}(t,E)=[E+i\eta-H_{\alpha}(t)]^{-1}.\label{eq-LFGr}
\end{split}
\end{equation}
Then Eq.~(\ref{eq-LGr-1}) becomes
\begin{equation}
\begin{split}
[g^r_{\alpha,f}(t,E)]^{-1}g^r_{\alpha}(t,E)+\frac{i}{2}\{(g^r_{\alpha,f})^{-1},g^r_{\alpha}\}=I.
\end{split}\label{eq-LGr-2}
\end{equation}

Since the Poisson bracket is actually calculating the first order term, $g^r_{\alpha}$ in it should be replaced by the Frozen Green's function $g^r_{\alpha,f}$ directly. And the term $g^r_{\alpha,f}\{(g^r_{\alpha,f})^{-1},g^r_{\alpha,f}\}$ equals to
\begin{equation}
\begin{split}
&g^r_{\alpha,f}\partial_t(g^r_{\alpha,f})^{-1}\partial_Eg^r_{\alpha,f}- g^r_{\alpha,f}\partial_E(g^r_{\alpha,f})^{-1}\partial_tg^r_{\alpha,f}\\
=&-g^r_{\alpha,f}\partial_tH_{\alpha}g^r_{\alpha,f}g^r_{\alpha,f}+g^r_{\alpha,f}g^r_{\alpha,f}\partial_tH_{\alpha}g^r_{\alpha,f}\\
=&~ g^r_{\alpha,f}(-\partial_tH_{\alpha}g^r_{\alpha,f}+g^r_{\alpha,f}\partial_tH_{\alpha})g^r_{\alpha,f}\\
=&~ 0,
\end{split}\label{eq22}
\end{equation}
where we have used the relation $-\partial_tH_{\alpha}g^r_{\alpha,f}+g^r_{\alpha,f}\partial_tH_{\alpha}=0$ that can be proved by using $\partial_t[(g^r_{\alpha,f})^{-1}\partial_tg^r_{\alpha,f}]=0$.

Therefore, from Eqs.~(\ref{eq-LGr-2}) and (\ref{eq22}), we have
\begin{equation}
\begin{split}
g^r_{\alpha}(t,E)=g^r_{\alpha,f}(t,E).\label{eq-LGr-3}
\end{split}
\end{equation}
Then the self energy $\Sigma^r_{\alpha}(t,E)$ is equal to
\begin{equation}
\begin{split}
\Sigma^r_{\alpha}(t,E)=H_{C\alpha}g^r_{\alpha,f}(t,E)H_{\alpha C}.\label{eq-SelfR-1}
\end{split}
\end{equation}

Furthermore, we calculate the lesser self energy up to the first order in time. As shown in Ref.~[\onlinecite{ydw2}], the lesser Green's function of lead $\alpha$ can be obtained from retarded and advanced Green's functions
\begin{equation}
\begin{split}
g^<_{k\alpha}(t_1,t_2)=if(\epsilon^{(0)}_{k\alpha})[g^r_{k\alpha}(t_1,t_2)-g^a_{k\alpha}(t_1,t_2)],\label{eq-LGl-1}
\end{split}
\end{equation}
where $g^{r,a}_{k\alpha}$ is defined as
\begin{eqnarray}
g^{r,a}_{k\alpha}(t,t')=\mp i\theta (\pm t \mp t')
\exp[-i\int^t_{t'} dt_1 \epsilon_{k\alpha} (t_1)].
\label{X2grlead1}
\end{eqnarray}

By taking the Wigner transformation, Eq.~(\ref{eq-LGl-1}) can be written as
\begin{equation}
\begin{split}
g^<_{k\alpha}(t,E)=if(\epsilon^{(0)}_{k\alpha})[g^r_{k\alpha}(t,E)-g^a_{k\alpha}(t,E)],\label{eq-LGl-2}
\end{split}
\end{equation}
where $g^{r/a}_{k\alpha}(t,E)=(E\pm i\eta-\epsilon_{k\alpha} (t))^{-1}$ due to Eqs.~(\ref{eq-LFGr}) and (\ref{eq-LGr-3}). According to the definition, the lesser self energy can be expressed as
\begin{equation}
\begin{split}
[\Sigma^<_{\alpha}(t,E)]_{mn}&=\sum_{k}t^*_{k\alpha m} g^<_{k\alpha}(t,E) t_{k\alpha n}\\
&=if_{\alpha}(E)[\Gamma_{\alpha}(E-qv_{\alpha}(t))]_{mn},\label{eq-LGl-2}
\end{split}
\end{equation}
where $f_\alpha(E,t) \equiv f(E-qv_\alpha(t))$ is the Fermi distribution function of lead $\alpha$ with applied bias. The linewidth function $\Gamma_\alpha$ is defined as
\begin{equation}
[\Gamma_\alpha(E)]_{mn} \equiv 2\pi \sum_k t^*_{k\alpha m} t_{k\alpha n} \delta(E-\epsilon^{(0)}_{k\alpha}). \label{X2gamma}
\end{equation}
It is also equal to
\begin{equation}
\begin{split}
\Gamma_\alpha(t,E)=i[\Sigma^r_\alpha(t,E)-\Sigma^a_\alpha(t,E)]. \label{eq-LGl-3}
\end{split}
\end{equation}

Therefore, the self energy in Eqs.~(\ref{eq-Gr-2}) and (\ref{eq-Gl-2}) can be calculated by using Frozen Green's function of the lead. Finally, we introduce the Frozen retarded Green's function of the central scattering region
\begin{equation}
G^r_f(t,E)=[G^a_f(t,E)]^{\dagger}=[E-H_C(t)-\Sigma^r(t,E)]^{-1}. \label{eq-fGr}
\end{equation}
Then Eqs.~(\ref{eq-Gr-2}) and (\ref{eq-Gl-2}) become
\begin{equation}
G^r(t,E)=G^r_{f}(t,E)+\frac{i}{2}G^r_{f}\{(G^r_{f})^{-1},G^r\},\label{eq-Gr-3}
\end{equation}
and
\begin{equation}
\begin{split}
G^<(t,E)=&G^r_{f}(t,E)\Sigma^<(t,E)G^{a}_f(t,E)\\
&+\frac{i}{2}G^r_{f}[\left\{\Sigma^r,G^<\right\}+\left\{\Sigma^<,G^a\right\}].\label{eq-Gl-3}
\end{split}
\end{equation}
Furthermore, we express Floquet Green's function in terms of Frozen Green's function up to the first order in time,
\begin{equation}
G^r(t,E)\simeq G^r_{f}(t,E)+g^r,\label{eq-Gr-4}
\end{equation}
and
\begin{equation}
\begin{split}
G^<(t,E)&\simeq G^r_{f}(t,E)\Sigma^<(t,E)G^{a}_f(t,E)+g^<\\
&=G^{<}_f + g^<,\label{eq-Gl-4}
\end{split}
\end{equation}
where we have introduced lesser Frozen Green's function $G^{<}_f=G^r_{f}(t,E)\Sigma^<(t,E)G^{a}_f(t,E)$. After some derivations, we have the following relations
\begin{equation}
\begin{split}
g^r = \frac{i}{2}\big[(\partial_E G^r_f)\partial_t(H_C+\Sigma^r)G^r_f - G^r_f\partial_t(H_C+\Sigma^r)(\partial_E G^r_f)\big],
\end{split}\label{eq-Gr-5}
\end{equation}
and
\begin{equation}
\begin{split}
g^{<} =& \frac{i}{2}\big[(\partial_E G^{<}_f)\partial_t(H_C+\Sigma^a)G^a_f - G^{<}_f\partial_t(H_C+\Sigma^a)(\partial_E G^a_f) \\
&+ (\partial_E G^{r}_f)(\dot{H}_C+\dot{\Sigma}^r)G^{<}_f - G^{r}_f(\dot{H}_C+\dot{\Sigma}^r)(\partial_E G^{<}_f) \\
&+ (\partial_E G^{r}_f)\dot{\Sigma}^{<}G^a_f - G^{r}_f\dot{\Sigma}^{<}(\partial_E G^a_f) \big], \label{eq-Gl-5}
\end{split}
\end{equation}
where the relation $\partial_{E/t}[(G^{r}_f)^{-1}G^r_f]=0$ is used.

Notice that the time-dependent Coulomb interaction $U(t)$ is obtained through solving the Poisson equation Eq.~(\ref{eq-Poisson-1}). The current conservation law is expressed as,
\begin{equation}
\sum_{\alpha}I_{\alpha}(t)+\partial_t \mathcal{Q}(t)=0, \label{eq-CurrentCon-1}
\end{equation}
with total charge of the system
\begin{equation}
\mathcal{Q}(t)=-i \int\frac{dE}{2\pi}{\rm Tr}[G^<(t,E)]. \label{eq-Q-1}
\end{equation}
Since we wish to expand the time-dependent current $I_{\alpha}(t)$ up to the first order in frequency, the first order of lesser Green's function $g^{<}$ can be safely neglected during calculation of $\mathcal{Q}(t)$, i.e.,
\begin{equation}
\begin{split}
\mathcal{Q}(t)=-i \int\frac{dE}{2\pi}{\rm Tr}[G^<_f(t,E)].\label{eq-Q-2}
\end{split}
\end{equation}
Thus the corresponding Poisson equation
\begin{equation}
\begin{split}
\nabla^2 U(x,t)= -4\pi i [G^<_f(t,t)]_{xx},\label{eq-Poisson-2}
\end{split}
\end{equation}
can be solved. Therefore, the Coulomb potential $U(t)$ and $G^<_f(t,t)$ should be solved self-consistently. In the following section, we will show that once the Coulomb potential is self-consistently obtained, current conservation $\sum_{\alpha}I_{\alpha}(t)=0$ is automatically guaranteed.

\section{Time-dependent current formula}
Similarly, the time-dependent current $I_{\alpha}(t)$ in Eq.~(\ref{current-1}) can also be expressed in Wigner representation
\begin{equation}
\begin{split}
I_{\alpha}(t) =  & \int\frac{dE}{2\pi} {\rm Tr}\big[G^r(t,E)\Sigma^{<}_{\alpha}(t,E)+ G^{<}(t,E)\Sigma^a_{\alpha}(t,E)\\
& - \Sigma^{<}_{\alpha}(t,E)G^a(t,E) - \Sigma^r_{\alpha}(t,E)G^{<}(t,E) \\ &
+\frac{i}{2}\{G^r,\Sigma^{<}_{\alpha}\} + \frac{i}{2}\{G^{<},\Sigma^a_{\alpha}\} \\
&- \frac{i}{2}\{\Sigma^{<}_{\alpha},G^a\} - \frac{i}{2}\{\Sigma^{r}_{\alpha},G^{<}\} \big],
\end{split}\label{current2}
\end{equation}
where $\{...\}$ is the Poisson bracket defined in Eq.~(\ref{bracket}). Green's functions and self energies inside the bracket is short for operators in Wigner representation, {\it e.g.}, $G^{r,a,<} \equiv G^{r,a,<}(t,E)$, etc. Note that the time-dependent Coulomb potential is explicitly included in the frozen Green's function through Eq.~(\ref{eq-fGr}) and hence the time-dependent current.


Based on Eqs.~(\ref{eq-Gr-4}) and (\ref{eq-Gl-4}), the current can be divided into two parts: $I_{\alpha} = I_{\alpha}^{(0)} + I_{\alpha}^{(1)}$. Here, $I_{\alpha}^{(0)}$ corresponds to the adiabatic current and $I_{\alpha}^{(1)}$ is the first order correction due to frequency. With the help of the relation $G^{<}_f=G^r_{f}(t,E)\Sigma^<(t,E)G^{a}_f(t,E)$, the adiabatic current $I_{\alpha}^{(0)}$ can be written as
\begin{equation}
\begin{split}
I_{\alpha}^{(0)}(t) &= \int\frac{dE}{2\pi}{\rm Tr}[G^r_f\Sigma^{<}_{\alpha} + G^{<}_f\Sigma^a_{\alpha} - \Sigma^{<}_{\alpha}G^a_f - \Sigma^r_{\alpha}G^{<}_f]\\
&=\int\frac{dE}{2\pi}{\rm Tr}[(G^r_f-G^a_f)\Sigma^{<}_{\alpha} + iG^{<}_f\Gamma_{\alpha}]\\
&=\sum_{\beta}\int\frac{dE}{2\pi}{\rm Tr}[\Gamma_{\alpha}G^r_f\Gamma_{\beta}G^a_f](f_{\alpha}-f_{\beta}),
\end{split}\label{current1-0}
\end{equation}
where $G^a_f-G^r_f=iG^r_f\Gamma G^a_f$ is used. Correspondingly, the first order correction current is
\begin{equation}
\begin{split}
I_{\alpha}^{(1)}(t) = \int\frac{dE}{2\pi}{\rm Tr}\big[ & g^r\Sigma^{<}_{\alpha} + g^{<}\Sigma^a_{\alpha} - \Sigma^{<}_{\alpha}g^a - \Sigma^r_{\alpha}g^{<}  \\
& + \frac{i}{2}\{G^r_f,\Sigma^{<}_{\alpha}\} + \frac{i}{2}\{G^{<}_f,\Sigma^a_{\alpha}\}\\
& - \frac{i}{2}\{\Sigma^{<}_{\alpha},G^a_f\} - \frac{i}{2}\{\Sigma^{r}_{\alpha},G^{<}_f\} \big].
\end{split}\label{current1-1}
\end{equation}
Plugging Eqs.~(\ref{eq-Gr-5}) and (\ref{eq-Gl-5}) into Eq.~(\ref{current1-1}), the explicit expression of $I_{\alpha}^{(1)}$ can be obtained after some straightforward algebra. Then we have
\begin{equation}
\begin{split}
I_{\alpha}^{(1)}(t) = &\int\frac{dE}{2\pi}{\rm Tr}\{ (g^r-g^a)\Sigma^{<}_{\alpha} + ig^{<}\Gamma_{\alpha}  \\
&+\frac{i}{2}[\partial_E(G^r_f+G^a_f)\partial_t{\Sigma}^{<}_{\alpha}-\partial_t(G^r_f+G^a_f)\partial_E{\Sigma}^{<}_{\alpha}\\
&+\partial_t(\Sigma^r_{\alpha}+\Sigma^a_{\alpha})\partial_E{G}^{<}_f-\partial_E(\Sigma^r_{\alpha}+\Sigma^a_{\alpha})\partial_t{G}^{<}_f] \big\}.
\end{split}
\label{appeq7}
\end{equation}
From $G^{<}_f = G^r_f\Sigma^{<}G^a_f$, we have $\partial_{E/t} G^{<}_f = (\partial_{E/t} G^r_f)\Sigma^{<}G^a_f + G^r_f(\partial_{E/t} \Sigma^{<})G^a_f + G^r_f\Sigma^{<}(\partial_{E/t}  G^a_f)$. Eq. (\ref{appeq7}) can be further expressed as
\begin{equation}
\begin{split}
I_{\alpha}^{(1)}(t) = \int\frac{dE}{2\pi}{\rm Tr}\{ & (g^r-g^a)\Sigma^{<}_{\alpha} + i(G^r_f\Sigma^{<}g^a+g^r\Sigma^{<}G^a_f)\Gamma_{\alpha}\\
&-\frac{1}{2}(\partial_t G^r_f\Sigma^{<}G^a_f-G^r_f\Sigma^{<}\partial_t G^a_f)\partial_E\Gamma_{\alpha}\\
&+\frac{1}{2}(\partial_E G^{r}_f\Sigma^{<}G^a_f - G^{r}_f\Sigma^{<}\partial_E G^a_f)\partial_t{\Gamma}_{\alpha}\\
&+\partial_t\frac{d\mathfrak{N}_{\alpha}}{dE} -\frac{1}{2}\mathcal{P}\Gamma_{\alpha}\big\},\label{current1-f}
\end{split}
\end{equation}
where $g^a\equiv(g^r)^{\dagger}$ and we have used integration by parts for energy and introduced two terms
\begin{eqnarray}
\begin{split}
\mathcal{P}\equiv\partial_tG^{r}_f\Sigma^{<}\partial_E G^a_f-\partial_E G^{r}_f\Sigma^{<}\partial_tG^a_f,
\end{split}
\end{eqnarray}
\begin{equation}
\begin{split}
\frac{d\mathfrak{N}_{\alpha}}{dE}\equiv\frac{i}{2}&[\partial_E(G^r_f+G^a_f)\Sigma^{<}_{\alpha}-\partial_E(\Sigma^r_{\alpha}+\Sigma^a_{\alpha}){G}^{<}_f\\
&+i(\partial_E G^{r}_f\Sigma^{<}G^a_f - G^{r}_f\Sigma^{<}\partial_E G^a_f)\Gamma_{\alpha}].
\end{split}
\label{appeq8}
\end{equation}
The physical meaning of $[d\mathfrak{N}_{\alpha}/dE]_{xx}$ is similar to injectivity\cite{buttiker1996,buttiker1997,Texier}: the local charge at position $x$ that an electron injects from lead $\alpha$ and emits to all leads due to the external bias.

Since current conservation is one of the basic requirements in quantum transport theory, we demonstrate that the current shown in Eq.~(\ref{current1-f}) is conserved when the Coulomb interaction is included. Firstly, it is straightforward to show the adiabatic current satisfies
\begin{equation}
\begin{split}
\sum_{\alpha}I_{\alpha}^{(0)} = \sum_{\alpha\beta}\int\frac{dE}{2\pi}{\rm Tr}[\Gamma_{\alpha}G^r_f\Gamma_{\beta}G^a_f](f_{\alpha}-f_{\beta}) = 0.
\end{split}
\end{equation}
Then from Eq.~(\ref{current1-f}), we have
\begin{equation}
\begin{split}
\sum_{\alpha}I_{\alpha}^{(1)} = \sum_{\alpha}\partial_t\frac{d\mathfrak{N}_{\alpha}}{dE}=-i\partial_t\int\frac{dE}{2\pi}{\rm Tr}[G^<_f]=-\partial_t\mathcal{Q}(t),
\end{split}
\label{appeq18}
\end{equation}
where total charge $\mathcal{Q}(t)$ is defined in Eq.~(\ref{eq-Q-2}). In the central scattering region, the Poisson equation in Eq.~(\ref{eq-Poisson-2}) should be solved with proper boundary conditions. The natural and reasonable conditions are zero electric fields on the boundaries leading to constant electric potential $U$ there. Thus we have ${\rm Tr}\left[\nabla^2U(x,t)\right]=0$ due to these boundary conditions of the Poisson equation. Physically, the total charge in the central region $\mathcal{Q}(t)$ is zero and so is its derivative, which is known as the charge neutrality condition. Then we arrive at $\sum_\alpha I_\alpha(t)=0$.

\subsection{Wide band limit}
Wide band limit (WBL) is usually adopted to simplify the analysis of transport properties. In this subsection, we will derive formulas for the time-dependent current under WBL, where self energies are approximately independent of both energy and time, i.e.,
\begin{equation}
\begin{split}
&\partial_E \Sigma^{r/a}=0;\partial_E \Gamma=0;\\
&\partial_t \Sigma^{r/a}=0;\partial_t \Gamma=0.
\end{split}\label{wbl-1}
\end{equation}
Therefore, we have
\begin{eqnarray}
\begin{split}
\partial_tG^{r/a}_f&=G^{r/a}_f\partial_tHG^{r/a}_f,\\
\partial_EG^{r/a}_f&=-G^{r/a}_fG^{r/a}_f,\\
G^<_f&=G^r_f(t,E)\Sigma^{<}G^a_f(t,E),\\
\Sigma^{<}&=\sum_{\alpha}i\Gamma_{\alpha}f_{\alpha}(E,t),
\end{split}\label{wbl-2}
\end{eqnarray}
and
\begin{equation}
\begin{split}
g^r &= \frac{i}{2}\big[\partial_t G^r_fG^r_f-G^r_f\partial_t G^r_f\big],\\
g^a &= \frac{i}{2}\big[\partial_t G^a_fG^a_f-G^a_f\partial_t G^a_f\big].
\end{split}\label{wbl-3}
\end{equation}
Note that the lesser self energy $\Sigma^<$ is still time-dependent due to the presence of Fermi distribution function $f_{\alpha}(E,t)$ in it. Under WBL, the zeroth order current $I_{\alpha}^{(0)}$ has the same form
\begin{equation}
\begin{split}
I_{\alpha}^{(0)}(t) =\sum_{\beta}\int\frac{dE}{2\pi}{\rm Tr}[\Gamma_{\alpha}G^r_f\Gamma_{\beta}G^a_f](f_{\alpha}(E,t)-f_{\beta}(E,t)),
\end{split}\label{current1-wbl-0}
\end{equation}
where $\Gamma_{\alpha}$ is energy-independent. By using Eqs.~(\ref{wbl-1}), (\ref{wbl-2}), and (\ref{wbl-3}), the first order current $I_{\alpha}^{(1)}(t)$ in Eq.~(\ref{current1-f}) can be greatly simplified as
\begin{equation}
\begin{split}
I_{\alpha}^{(1)}(t) =&\int\frac{dE}{2\pi}{\rm Tr}[-i(G^r_f \partial_t G^r_f+\partial_t G^a_f G^a_f)\Sigma^<_{\alpha}\\
&+(G^r_f\partial_tG^r_f\Sigma^<G^a_f - G^r_f \Sigma^<\partial_t G^a_f G^a_f)\Gamma_{\alpha}\\
&+\frac{1}{2}(-i(G^r_f G^r_f + G^a_f G^a_f)\partial_t \Sigma^{<}_{\alpha}\\
&-(-G^r_f G^r_f \partial_t \Sigma^{<}G^a_f + G^{r}_f \partial_t \Sigma^{<} G^a_f G^a_f)\Gamma_{\alpha})].
\end{split}\label{current-wbl-2}
\end{equation}

We can also check the validity of current conservation for the zeroth and first order currents in Eqs.~(\ref{current1-wbl-0}) and (\ref{current-wbl-2}). The current conservation of zeroth order can be easily obtained, and the summation of the first order currents in Eq.~(\ref{current-wbl-2}) gives
\begin{equation}
\begin{split}
\sum_{\alpha}I_{\alpha}^{(1)}=&\int\frac{dE}{2\pi}{\rm Tr}[-i(G^r_f\partial_tG^r_f+\partial_tG^a_fG^a_f)\Sigma^<\\
&+(G^r_f\partial_tG^r_f\Sigma^<G^a_f-G^r_f\Sigma^<\partial_tG^a_fG^a_f)\Gamma\\
&+\frac{1}{2}(-i(G^r_fG^r_f+G^a_fG^a_f)\partial_t \Sigma^{<}\\
&-(-G^r_fG^r_f\partial_t \Sigma^{<}G^a_f + G^{r}_f\partial_t \Sigma^{<}G^a_fG^a_f)\Gamma)]\\
=&-i\int\frac{dE}{2\pi}{\rm Tr}[\partial_tG^r_f\Sigma^<G^a_f \\
 &+G^r_f\partial_t\Sigma^<G^a_f+G^r_f\Sigma^<\partial_tG^a_f]\\
=&-i\partial_t \int\frac{dE}{2\pi}{\rm Tr} [G^<_f] \\
=& -\partial_t \mathcal{Q}(t)=0.
\end{split}\label{current-wbl-3}
\end{equation}
To summarize, the time-dependent current expressed in Eq. (\ref{current-wbl-2}) under WBL is also conserved.

\subsection{Small ac bias limit}
At the small ac bias limit, we can derive analytic expressions of the dynamic conductance and time-dependent currents, which gives intuitive understanding on quantum ac transport properties. Meanwhile, these analytic results can be compared with the previous work Ref.~[\onlinecite{ydw2}] which focused on quantum ac transport theory at finite frequency and small bias limit.

In general, time-dependent currents in Eqs.~(\ref{current1-wbl-0}) and (\ref{current-wbl-2}) are in nonlinear response to bias voltage. In the following, we expand the time-dependent current $I_{\alpha}(t)$ order by order in the bias voltage. First, we need to separate the Hamiltonian into two parts: $H_C(t)=H_{C0}+U(t)$, where $H_{C0}$ is time-independent and $U(t)$ is the time-dependent Coulomb potential. $U(t)$ can be expanded in terms of the bias magnitude $v_{\alpha}(t=0)=v_{\alpha}$\cite{ydw2},
\begin{equation}
\begin{split}
U(t) &= U_{eq} + U_{1}(t) + U_{2}(t) + ... \\
&= U_{eq} + q\sum_{\alpha}u_{\alpha}(t)v_{\alpha} + \frac{1}{2}q^2\sum_{\alpha\beta}u_{\alpha\beta}(t)v_{\alpha}v_{\beta}+...,
\end{split}\label{U-1}
\end{equation}
with $U_{eq}$ the equilibrium potential in the absence of external bias. $U_1(t)$ and $U_2(t)$ are the first and second order corrections due to the presence of external bias, respectively. Correspondingly, $u_{\alpha}(t)$ and $u_{\alpha\beta}(t)$ are the first and second order characteristic potentials\cite{buttiker0,ydw2,zsm}. Under Thomas-Fermi approximation, the Poisson-like equations for these characteristic potentials are defined as
\begin{eqnarray}\label{U-2}
&& -\nabla^2 u_{\alpha}(t,x) + 4\pi \frac{dn(t,x)}{dE} u_{\alpha}(t,x) = 4\pi \frac{dn_\alpha(t,x)}{dE},\nonumber \\
&& -\nabla^2 u_{\alpha \beta}(t,x) +4\pi \frac{dn}{dE} u_{\alpha \beta}(t,x) = 4\pi \frac{d\tilde{n}_{\alpha \beta}(t,x)}{dE},
\end{eqnarray}
where
\begin{eqnarray}
\frac{d\tilde{n}_{\alpha \beta}}{dE} = \frac{d^2n_\alpha}{dE^2} \delta_{\alpha \beta} -\frac{d^2n_\alpha}{dE^2} u_\beta -
\frac{d^2n_\beta}{dE^2} u_\alpha +\frac{d^2n}{dE^2} u_\alpha u_\beta. \nonumber
\end{eqnarray}
Here $dn_{\alpha}/dE$ is the time-dependent injectivity, and $d^2n_\alpha/dE^2 = \partial_E dn_\alpha/dE$. The procedure for calculating the first and second order characteristic potentials is shown in Appendix B.

For simplicity, WBL is adopted in the following discussion. The Frozen Green's function in Eq.~(\ref{eq-fGr}) can be expressed as
\begin{equation}
\begin{split}
G^r_f(t,E)&=[E-H'_{C0}-U_{eq}-\sum_n U_n(t)-\Sigma^r]^{-1}\\
&=G^r_0+G^r_0\sum_nU_nG^r_f,
\end{split}\label{eq-fGr-wbl-1}
\end{equation}
where equilibrium Green's function is defined as $G^r_0\equiv[E-H'_{C0}-U_{eq}-\Sigma^r]^{-1}$ and self energy $\Sigma^r$ is independent of energy; $H'_{C0}=H_{C0}-U_{eq}$ and $U_n$ is the $n$-th order correction to the time-dependent Coulomb potential.

Now we expand the nonequilibrium Green's function in terms of the first order bias. Thus the Frozen Green's function in Eq.~(\ref{eq-fGr-wbl-1}) can be further written as
\begin{equation}
\begin{split}
G^r_f(t,E)&=G^r_0+G^r_0U_1(t)G^r_0.
\end{split}\label{eq-fGr-wbl-2}
\end{equation}
Correspondingly, the zeroth order current $I_{\alpha}^{(0)}$(t) in Eq.~(\ref{current1-wbl-0}) is given by
\begin{equation}
\begin{split}
I_{\alpha}^{(0)}(t) &=\sum_{\beta}\int\frac{dE}{2\pi}{\rm Tr}[\Gamma_{\alpha}G^r_0\Gamma_{\beta}G^a_0](\partial_E f_0)(v_{\alpha}(t)-v_{\beta}(t)).
\end{split}\label{current1-wbl-v-1}
\end{equation}
Here we have used the Taylor expansion of Fermi function $f_{\alpha}=f_0+(\partial_E f_0)v_{\alpha}(t)$ up to the first order in bias and $f_0$ is the equilibrium Fermi distribution. In general, by taking Fourier transformation, we can obtain the frequency-dependent current $I_{\alpha}({\Omega})$, where $\Omega$ is the response frequency. From the definition of dynamic conductance $I_{\alpha}({\Omega})=\sum_{\beta}G_{\alpha\beta}(\Omega)v_{\beta}(\Omega)$ and $v_{\beta}(\Omega)=\pi v_{\alpha}[\delta(\Omega+\omega)+\delta(\Omega-\omega)]$, we find dynamic conductance $G^{(0)}_{\alpha\beta}(0)$ of the zeroth order current $I_{\alpha}^{(0)}(t)$,
\begin{equation}
\begin{split}
G^{(0)}_{\alpha\beta}(0)=\sum_{\beta}\int\frac{dE}{2\pi}{\rm Tr}[\Gamma_{\alpha}G^r_0\Gamma G^a_0\delta_{\alpha\beta}+\Gamma_{\alpha}G^r_0\Gamma_{\beta}G^a_0](\partial_E f_0).
\end{split}\label{dy1-cond-2}
\end{equation}
This result agrees with Eq.~(35) of Ref.~[\onlinecite{ydw2}] when $\Omega=0$.

Furthermore, we can calculate the first order dynamic conductance $G^{(1)}_{\alpha\beta}(\Omega)$. Since only the first order bias voltage is considered, Eq. (\ref{current-wbl-2}) is simplified as
\begin{equation}
\begin{split}
I_{\alpha}^{(1)}(t)& =\int\frac{dE}{2\pi}{\rm Tr}[(G^r_0G^r_0\partial_tU_1G^r_0+G^a_0\partial_tU_1G^a_0G^a_0)\Gamma_{\alpha}f_0\\
&+i(G^r_0G^r_0\partial_tU_1G^r_0\Gamma G^a_0-G^r_0\Gamma G^a_0\partial_tU_1G^a_0G^a_0)\Gamma_{\alpha}f_0\\
&+\frac{1}{2}((G^r_0G^r_0+G^a_0G^a_0)\Gamma_{\alpha}\partial_tf_{\alpha}\\
&-i\sum_{\beta}(-G^r_0G^r_0\Gamma_{\beta} G^a_0 + G^{r}_0\Gamma_{\beta}G^a_0G^a_0)\Gamma_{\alpha}\partial_tf_{\alpha}]\\
=&\int\frac{dE}{2\pi}{\rm Tr}[(G^r_0G^r_0\partial_tU_1G^a_0+G^r_0\partial_tU_1G^a_0G^a_0)\Gamma_{\alpha}f_0\\
&+\frac{1}{2}((G^r_0G^r_0+G^a_0G^a_0)\Gamma_{\alpha}\partial_Ef_0\partial_t v_{\alpha}(t)\\
&-i\sum_{\beta}(-G^r_0G^r_0\Gamma_{\beta} G^a_0 + G^{r}_0\Gamma_{\beta}G^a_0G^a_0)\Gamma_{\alpha}\partial_Ef_0\partial_t v_{\beta}(t))],
\end{split}\label{dy2-cond-1}
\end{equation}
where $\partial_t f_{\alpha}(E,t)=\partial_E f_{\alpha}(E,t) \partial_t v_{\alpha}(t)$ and $iG^r_0\Gamma G^a_0=G^a_0-G^r_0$. Fourier transformation of Eq. (\ref{dy2-cond-1}) gives
\begin{equation}
\begin{split}
I_{\alpha}^{(1)}(\Omega) =&-i\Omega\int\frac{dE}{2\pi}{\rm Tr}[(G^r_0G^r_0U_1(\Omega)G^a_0+G^r_0U_1(\Omega)G^a_0G^a_0)\Gamma_{\alpha}f_0\\
&+\frac{1}{2}((G^r_0G^r_0+G^a_0G^a_0)\Gamma_{\alpha}\partial_Ef_0v_{\alpha}(\Omega)\\
&-i\sum_{\beta}(-G^r_0G^r_0\Gamma_{\beta} G^a_0 + G^{r}_0\Gamma_{\beta}G^a_0G^a_0)\Gamma_{\alpha}\partial_Ef_0v_{\beta}(\Omega))]\\
=&i\Omega\int\frac{dE}{2\pi}(- \partial_E f_0)\sum_{\beta}{\rm Tr}[-G^r_0u_{\beta}(\Omega)G^a_0\Gamma_{\alpha}\\
&+\frac{1}{2}((G^r_0G^r_0+G^a_0G^a_0)\Gamma_{\alpha}\delta_{\alpha\beta}\\
&-i(-G^r_0G^r_0\Gamma_{\beta} G^a_0 + G^{r}_0\Gamma_{\beta}G^a_0G^a_0)\Gamma_{\alpha})]v_{\beta}(\Omega).\\
\end{split}\label{dy2-cond-2}
\end{equation}
Thus the first order dynamic conductance $G^{(1)}_{\alpha\beta}(\Omega)$ is
\begin{equation}
\begin{split}
G^{(1)}_{\alpha\beta} =&i\Omega\int\frac{dE}{2\pi}(- \partial_E f_0){\rm Tr}[\frac{1}{2}(G^r_0G^r_0+G^a_0G^a_0)\Gamma_{\alpha}\delta_{\alpha\beta}\\
&+\frac{i}{2}(G^r_0G^r_0\Gamma_{\beta} G^a_0 - G^{r}_0\Gamma_{\beta}G^a_0G^a_0)\Gamma_{\alpha}\\
&-G^r_0u_{\beta}(\Omega)G^a_0\Gamma_{\alpha}],
\end{split}\label{dy2-cond-2}
\end{equation}
which is exactly the same as that in Ref.~[\onlinecite{ydw2}]. Notice that the first order dynamic conductance is actually the emittance describing low frequency response of the system\cite{buttiker0}.

Here the consistency between Ref.~[\onlinecite{ydw2}] and the present work at the small ac bias limit is demonstrated. It is worth mentioning that, in Ref.~[\onlinecite{ydw2}], the Coulomb potential is treated in a perturbative way, which is suitable for describing ac transport at finite frequency and small ac bias limit. If one tries to simulate the finite ac bias case with the theory developed in Ref.~[\onlinecite{ydw2}], the Coulomb potential has to be expanded in terms of the characteristic potential order by order, which is very complicated especially for nonlinear terms. In contrast, the present work can deal with the Coulomb potential directly by solving the Poisson equation Eq.~(\ref{eq-Poisson-2}). Compared with Ref.~[\onlinecite{ydw2}], the present work has advantages in investigating nonlinear ac transport with finite ac bias voltages at the low frequency limit. Ref.~[\onlinecite{ydw2}] and the present work focus on different regimes of quantum ac transport and complimentary to each other. The application of our quantum nonlinear ac theory to nonlinear electrochemical capacitance is shown below.

\subsection{C-V curve and nonlinear electrochemical capacitance}

In this subsection, we discuss the nonlinear emittance in a system where dc transport is not allowed, for example, the magnetic tunneling junction (MTJ). Then the nonlinear emittance is equivalent to the electrochemical capacitance\cite{buttiker93,jian-ec,xu14,Flindt16}.

One the other hand, we could assume that the scattering region of a two-probe system can be roughly divided into two regions, i.e., $\Omega_I$ and $\Omega_{II}$, and expand total charge of the system in terms of bias voltage
\begin{equation}
Q_{\alpha}(t)= q \sum_{\beta}C_{\alpha\beta}(t)v_{\beta}+q^2 \sum_{\beta\gamma}C_{\alpha\beta\gamma}(t)v_{\beta}v_{\gamma}+\cdots ,\label{ec-eq-1}
\end{equation}
where $\alpha = I,II$. Since the bias-dependent $Q_{\alpha}$ can be calculated though expanding Frozen lesser's Green's function $G^{<}_f$ in terms of $v_{\beta}(t)$, we have
\begin{equation}
\begin{split}
Q_{\alpha}(t) &=  -i \int_{\Omega_{\alpha}} dx ~[(G^r_f\Sigma^{<}G^a_f)]_{xx}\\
=  \sum_{\beta}&\int_{\Omega_{\alpha}} dx ~ [(G^r_0+G^r_0(U_1(t)+U_2(t))(G^r_0+G^r_0U_1(t)G^r_0)) \\
&\times \Gamma_{\beta}[f_0+\partial_Ef_0v_{\beta}(t)+\frac{1}{2}\partial^{2}_E f_0 v^2_{\beta}(t)] \\
&\times (G^a_0+G^a_0(U_1(t)+U_2(t))(G^a_0+G^a_0U_1(t)G^a_0))]_{xx},\label{ec-eq-2}
\end{split}
\end{equation}
where Eq.~($\ref{eq-fGr-wbl-1}$) is used. Finally, the linear capacitance $C_{\alpha\beta}(t)$ and nonlinear capacitance $C_{\alpha\beta\gamma}(t)$ are equal to
\begin{equation}
\begin{split}
C_{\alpha\beta}(t)=&\int_{\Omega_{\alpha}} dx ~ [\partial_Ef_0 G^r_0\Gamma_{\beta}G^a_0~cos(\omega t)\\
& - G^r_0(u_{\beta}(t)G^r_0\Gamma_{\beta}+\Gamma_{\beta}G^a_0u_{\beta}(t))G^a_0f_0]_{xx},\label{ec-eq-3}
\end{split}
\end{equation}
and
\begin{equation}
\begin{split}
C_{\alpha\beta\gamma}(t)=&\int_{\Omega_{\alpha}} dx ~ [\frac{1}{2}\partial^2_Ef_0 G^r_0\Gamma_{\beta}G^a_0\delta_{\beta\gamma}~cos^2(\omega t)\\
&-G^r_0(u_{\beta}(t)G^r_0\Gamma_{\gamma}+\Gamma_{\gamma}G^a_0u_{\beta}(t))G^a_0\partial_E f_0~cos(\omega t)\\
&-G^r_0(u_{\beta\gamma}(t)G^r_0\Gamma+\Gamma G^a_0u_{\beta\gamma}(t))G^a_0 f_0\\
&+G^r_0(u_{\beta}(t)G^r_0u_{\gamma}(t)G^r_0\Gamma \\
&+G^r_0\Gamma G^a_0u_{\beta}(t)G^a_0u_{\gamma}(t))G^a_0f_0]_{xx}.\label{ec-eq-4}
\end{split}
\end{equation}
Here WBL is not assumed in deriving Eq.~(\ref{ec-eq-4}). After obtaining the characteristic potentials $u_{\beta}(t)$ and $u_{\beta\gamma}(t)$, one can calculate the voltage-dependent linear and nonlinear electrochemical capacitances.

First principles calculation of the time-dependent current or electrochemical capacitance proceeds as follows. First, at a given time $t$, the lead Hamiltonians and self-energies are obtained from conventional DFT. Second, one needs to solve the Poisson equation Eq. (\ref{eq-Poisson-2}) self-consistently within NEGF-DFT. Once the self-consistent accuracy is reached, we obtain the nonequilibrium Hamiltonian $H_C(t)$ and the retarded and lesser Green's functions defined in Eqs.~(\ref{eq-Gr-4}) and (\ref{eq-Gl-4}). Finally, we can calculate the time-dependent current from Eqs.~(\ref{current1-0}) and (\ref{current1-1}) as well as the linear and nonlinear electrochemical capacitances from Eqs.~(\ref{ec-eq-3}) and (\ref{ec-eq-4}) to investigate quantum ac transport properties.

\section{Summary}

In summary, we have developed a quantum nonlinear ac transport theory based on NEGF. At low frequency, through expanding the time-dependent NEGF in terms of Frozen Green's function, we can explicitly include the time-dependent self-consistent Coulomb interaction in the time-dependent current, which is nonlinear in bias voltage. Therefore, this nonlinear response theory automatically satisfies both current conservation and gauge invariance conditions. More importantly, it can be directly combined with DFT to calculate ac transport properties from atomic first principles. As a demonstration, we discuss how to calculate the ac current and nonlinear electrochemical capacitance versus the bias voltage. This quantum nonlinear ac transport theory can be easily extended to multiterminal systems for describing nonlinear Hall responses at low frequency. Furthermore, inelastic process, such as the electron-phonon/photon interaction, can also be included in the present theoretical formalism as effective self-energies to discuss their effects in quantum ac transport.

$${\bf ACKNOWLEDGMENTS}$$

We acknowledge support from the National Natural Science Foundation of China (Grants No. 12074230, No. 12174262, No. 12034014, and No. 12174023). L. Zhang thanks the Fund for Shanxi ``1331 Project", the Shanxi Province 100-Plan Talent Program, Research Project Supported by Shanxi Scholarship Council of China.

\section*{Appendix A}
In this appendix, we present the derivation of Eq.~(\ref{eq-Gr-2}) in Wigner representation. Starting from Eq.~(\ref{eq-Gr-1}), we can first calculate the time derivative by changing variables into Wigner representation
\begin{equation}
\begin{split}
i\frac{\partial}{\partial t}G^r(t_1,t_2)&=\frac{i}{2}\frac{\partial}{\partial t}G^r(t,\tau)+i\frac{\partial}{\partial \tau}G^r(t,\tau).\label{eq-A-1}
\end{split}
\end{equation}
Then by taking Fourier transform ($\int d\tau e^{iE\tau}$) of each term in Eq.~(\ref{eq-Gr-1}) and using Eqs.~(\ref{eqAB1}) and (\ref{eqAB-1}), we have
\begin{equation}
\begin{split}
i\frac{\partial}{\partial \tau}G^r(t,\tau)&\rightarrow EG^r(t,E),\\
H_C(t)G^r(t,\tau)&\rightarrow H_C(t)G^r(t,E)-\frac{i}{2}\partial_tH_C(t)\partial_EG^r(t,E),\\
\int\Sigma^r(t,t_1)G^r(t_1,t')dt_1&\rightarrow \Sigma^r(t,E)G^r(t,E) + \frac{i}{2}\left\{\Sigma^r,G^r\right\}.\label{eq-A-Gr-1}
\end{split}
\end{equation}
Thus, we can easily arrive at Eq.~(\ref{eq-Gr-2}) and obtain the lesser Green's function defined in Eq.~(\ref{eq-Gl-2}).

\section*{ Appendix B }

In this appendix, we demonstrate how to calculate the characteristic potentials defined in Eq.~(\ref{U-2}). We first solve the first order potential. From Eq.~(\ref{U-2}), we have
\begin{equation}
\begin{split}
-\nabla^2 u_{\alpha}(t,x) = 4\pi \frac{dn_\alpha(t,x)}{dE}-4\pi \frac{dn(t,x)}{dE} u_{\alpha}(t,x), \label{pe1}
\end{split}
\end{equation}
where the injectivity is defined as\cite{buttiker0,ydw2}
\begin{eqnarray}
\frac{dn_{\alpha}(t,x)}{dE} = -~cos(\omega t) \int_E \partial_E f [G^r_0 \Gamma_{\alpha} G^a_0]_{xx},
\end{eqnarray}
and $dn_{\alpha}(t,x)/dE$ satisfies
\begin{equation}
\sum_\alpha \frac{dn_{\alpha}(t,x)}{dE} = \frac{dn(t,x)}{dE}, \label{X2inj1}
\end{equation}
with $dn(t,x)/dE$ the local density of states at time $t$. In general, the Poisson equation Eq.~(\ref{pe1}) needs to be solved within the NEGF-DFT framework and $u_\alpha$ is self-consistently obtained. For the simplest case, we can adopt the quasineutrality approximation\cite{buttiker0}, which means that the charge density at each point is zero so that $-\nabla^2 u_{\alpha}(t,x) = 0$. Then analytic expression of the first order characteristic potential is shown as
\begin{equation}
u_\alpha = \frac{dn_\alpha(t,x)}{dE}/\frac{dn(t,x)}{dE}. \label{u1}
\end{equation}

When the first order potential $u_\alpha$ is obtained, we can calculate the second order injectivity $d\tilde{n}_{\alpha \beta}/dE$:
\begin{equation}
\begin{split}
\frac{d\tilde{n}_{\alpha \beta}}{dE} = \frac{d^2n_\alpha}{dE^2} \delta_{\alpha \beta} -\frac{d^2n_\alpha}{dE^2} u_\beta -
\frac{d^2n_\beta}{dE^2} u_\alpha +\frac{d^2n}{dE^2} u_\alpha u_\beta,
\end{split}\label{soinject}
\end{equation}
where the second order derivatives with respect to energy can be numerically facilitated with the finite difference method. Then substituting it into Eq.~(\ref{U-2}), we have
\begin{equation}
\begin{split}
-\nabla^2 u_{\alpha \beta}(t,x) = 4\pi \frac{d\tilde{n}_{\alpha \beta}(t,x)}{dE}- 4\pi \frac{dn}{dE} u_{\alpha \beta}(t,x),
\end{split}\label{pe2}
\end{equation}
which also needs to be solved. Notice that here $dn/dE = \sum_\alpha dn_\alpha/dE$ is the total density of states. Again we can use the quasineutrality approximation, which leads to
\begin{equation}
u_{\alpha \beta} = \frac{d\tilde{n}_{\alpha \beta}(t,x)}{dE}/\frac{dn}{dE}. \label{u1}
\end{equation}
At the small ac bias limit, the second order potential $u_{\alpha \beta}$ contributes to ac transport properties less than the first order one, and hence the quasineutrality approximation is appropriate here.

\bigskip

\noindent{$^{\dagger)}$These authors contributed equally to this work.}

\noindent{$^{*}$ jianwang@hku.hk}

\end{document}